\documentclass[preprint]{aastex}
\usepackage{psfig}

\begin{document}

\slugcomment{To appear in AJ}

\title{\sc A Search for {\rm\sc Hi} in E+A Galaxies}

\author{Tzu-Ching Chang\altaffilmark{1}, Jacqueline H. van Gorkom\altaffilmark{1},
Ann. I. Zabludoff\altaffilmark{2}, Dennis Zaritsky\altaffilmark{2}, and J. Christopher Mihos\altaffilmark{3}}

\altaffiltext{1}{Department of Astronomy, Columbia University, 550 West 120th
Street, New York, NY 10027; tchang@astro.columbia.edu, jvangork@astro.columbia.edu.}
\altaffiltext{2}{Steward Observatory, University of Arizona, Tucson, AZ 85721;
azabludoff@as.arizona.edu, dzaritsky@as.arizona.edu}
\altaffiltext{3}{Department of Astronomy, Case Western Reserve University, 10900
Euclid Avenue, Cleveland, OH 44106; hos@burro.astr.cwru.edu}

\begin{abstract}

We present the results of {\rm\sc Hi} line and radio continuum
observations of five nearby E+A galaxies. These galaxies have spectra
that are dominated by a young stellar component but lack the emission
lines characteristic of significant, on-going star formation.  They
are selected from a unique sample of 21 E+A's identified by
\citet{zab96} in their spectroscopic search for E+A galaxies using the
Las Campanas Redshift Survey, where over 11,000 nearby galaxies were
examined. The five E+A galaxies span a range of environments: three
are in the field and two are in clusters. Only one system was detected
in {\rm\sc Hi} emission, the field E+A galaxy EA1, with a total flux
of 0.30 $\pm$ 0.02 Jy km $s^{-1}$ and an {\rm\sc Hi} mass of $3.5 \pm
0.2 \times 10^{9} ~h^{-2}$ ~M$_{\odot}$.  The {\rm\sc Hi} morphology
and kinematics of EA 1 suggest a galaxy-galaxy interaction, with a
dynamical age of $\sim$ 6 $\times$ $10^{8} ~h^{-1}$ yr inferred from the
{\rm\sc Hi} tail lengths and velocities.  This age estimate is
consistent with the interpretation drawn from optical spectroscopy
that starbursts in E+A galaxies began (and subsequently
ended) within the last 10$^9$ yr.  Our {\rm\sc Hi} detection
limits are such that if the other E+A's in our sample had the {\rm\sc
Hi} properties of EA 1, we would have detected (or marginally
detected) them.  We conclude that
E+A galaxies have a range of {\rm\sc Hi} properties.  None of the
galaxies were detected in radio continuum emission, with upper limits
to the radio power of $\sim$ $10^{21}$ $h^{-2}$ W Hz$^{-1}$. Our
limits exclude the possibility that these E+A's are dust-enshrouded
massive starburst galaxies, but are insufficient to exclude modest
star formation rates of less than a few $h^{-2}$ M$_{\odot}$
yr$^{-1}$.

\end{abstract}


\section{INTRODUCTION}

In their seminal photometric study of rich clusters of galaxies at
redshifts $\leq$ 0.5, Butcher and Oemler found a significant excess of
blue, starforming galaxies in distant clusters relative to nearby
clusters \citep{but78,but84} and provided one of the
clearest pieces of evidence for galaxy evolution.  Although the cause
for the relatively recent decrease in the mean star formation rates of
cluster galaxies over the past few billion years is still not well understood,
spectroscopic studies of these galaxies reveal a further
puzzle. Sixty percent of these blue, supposedly starforming galaxies
have strong Balmer absorption lines with little or no evidence for
ongoing star formation from optical emission lines \citep{cou87}.  Because 
their spectra are a superposition of a young stellar population
characterized by A stars and an older population characterized by K
stars, typical of elliptical galaxies, these galaxies were termed E+A
galaxies\footnote{We use the term E+A for historical reasons.  Because
the first detections of these galaxies were
in distant clusters \citep{dre83}, ``E'' was used to 
represent the assumed morphology and ``A'' the dominant stellar
lines of the spectrum.  A better, purely spectroscopic designation would
be K+A \citep[c.f.][]{fra93}.}.

E+A galaxies are believed to be post-starburst galaxies, in which star
formation has been almost entirely quenched. Why star formation has
ceased is unknown, although most speculate that the gas reservoir has
been depleted either through highly efficient consumption or by
external forces that removed the gas, such as ram pressure stripping.
The purpose of this study is to detect and measure the gaseous content
and morphology of E+A galaxies in an attempt to help us understand why
star formation has ceased so suddenly in galaxies that were vigorously
forming stars only $\sim$ 1 Gyr ago.

The population of E+A galaxies shows a dramatic change as a function
of redshift. \citet{dre90} found in seven clusters at 0.35
$< z <$ 0.45 that E+A galaxies constitute a significant fraction of
the detected cluster members (23/172), while \citet{bel95}
found in the distant cluster Cl 0939+472 (z $\sim 0.41$) that 35 of
the 169 member objects ($21 \pm 7$ \%) are E+A galaxies.  These
results have been strengthened by more recent surveys probing larger
numbers of intermediate redshift clusters \citep[e.g.][]{pog99}.
On the other hand, at low redshift the fraction of E+A galaxies in
clusters is close to zero. \citet{fab91} find that E+A
galaxies form less than 1$\%$ of the population in nearby clusters.

So far, the E+A phenomenon has mostly been studied in cluster
environments. The galaxy statistics of intermediate-redshift field
surveys are much poorer than for clusters.
The first environmentally unbiased
search for E+A galaxies in the local universe was done by
\citet[hereafter Z96]{zab96}.  From 11,113 galaxies at 0.05 $< z <$ 0.13
in the Las Campanas Redshift Survey \citep[LCRS;][]{she96}, Z96
identified 21 well-defined nearby E+A galaxies, which have the
strongest Balmer absorption lines (the average of the equivalent
widths of $\langle H\beta\gamma\delta\rangle >$ 5.5 \AA\ ) and weakest
[O {\rm\sc ii}] emission-line equivalent widths ($<$ 2.5 \AA\ ) of all
the galaxies in the survey. The fraction of nearby E+A galaxies in the
LCRS sample of Z96 is less than $0.2\%$. However, the comparison to
the higher redshift results is not straightforward because although
the general spectral characteristics --- strong Balmer absorption and
no [O {\rm\sc ii}] emission --- are similar for distant E+A's and the
Z96 sample, the latter sample had more stringent selection
criteria.  Interestingly 75$\%$ of the E+A's in the Z96 sample lie in
the field, well outside of clusters and rich groups of galaxies.

Although most scenarios explaining the E+A phenomenon have
concentrated on star formation triggers specific to the cluster
environment, the finding by Z96 that 75$\%$ of the local E+A sample
are in the field indicates that either there are different triggers
for cluster and field E+A's or the trigger does not depend on global
environment.  Z96 argue that {\it if} only one mechanism is responsible for the
E+A formation, then galaxy-galaxy interactions and mergers are that
mechanism: optical images of at least five of the 21 galaxies in their sample show
tidal features.

Finally, work by \citet{sma99} has cast some doubt on the
post-starburst interpretation of E+A galaxies.  From their radio
continuum detections of E+A galaxies in a cluster at $z=0.41$, they
suggest that these galaxies are not post-starburst galaxies, but 
actually host highly obscured starbursts.

{\rm\sc Hi} imaging can address a number of questions pertaining to
the E+A phenomenon. If E+A galaxies are post-starburst
galaxies, what happened to their gas --- did the burst consume it all?
A comparison of the gas content and surface density distribution
between E+A's and normal star forming galaxies might provide clues as
to why E+A's are now quiescent. If galaxy interactions are indeed
responsible for triggering the original bursts, the {\rm\sc Hi}
morphology and kinematics might present a tidal signature 
even in systems with undisturbed optical morphologies \citep[e.g., the M81/M82/NGC3077
system;][]{yun92}.  Could there be significant star formation
in E+A's that is obscured by dust \citep[c.f.][]{sma99}? Radio
continuum measurements, obtained as a byproduct of the {\rm\sc Hi}
observations, can set useful limits on such a scenario. 
Could the local environment of E+A's play a secondary role, as a source for gas
for subsequent star formation? As 
another byproduct of the observations, we obtain an estimate of the
{\rm\sc Hi} content of neighboring galaxies and can place a limit on the
total local gas reservoir for star formation.

In this paper, we present the results of a search for neutral hydrogen
gas from a spectroscopically well-defined sample of nearby E+A
galaxies. Five E+A galaxies, EA 1, EA 2, EA 3, EA 4, and EA 11, are
drawn from the 21 E+A galaxies defined in the survey of Z96.
Four of these galaxies (EA 1-4) are selected because 1) they have been observed 
already with the Hubble Space Telescope Wide Field/Planetary Camera 2
(Zabludoff et al., in prep.) and 2) they are easily accessible with the VLA.
The fifth galaxy, EA 11, is included in our sample because, like EA 4, it 
lies in a known cluster environment (Z96) and thereby provides a comparison
with the three field E+A's.
The numerical suffix in the galaxy name ranks the
strength of the 4000 \AA\ break, $D_{4000}$, in the galaxy, which is a
measure of the composition of the stellar population of a galaxy.  
Galaxies, such as EA 1, with smaller $D_{4000}$ values are those most dominated
by the A-stellar component and are thus bluer in color.
The target galaxies lie at redshifts between 0.08 and 0.12.
Our paper is organized as follows: Section 2
contains a summary of the observations and a brief description of the
data reduction procedures. The {\rm\sc Hi} and the radio continuum
properties of the galaxies are presented in Section 3. In Section 4,
we discuss the implications of our results. We present our conclusions
in Section 5.  Throughout this paper we use $H_{0} = 100 ~h$ km
sec$^{-1}$ Mpc$^{-1}$ and $q_{0} = 0.5$.

\section{OBSERVATIONS AND DATA REDUCTION}

We observed the five E+A galaxies in our sample using NRAO's Very
Large Array (VLA) in January and March 1996 and June 1997.  Table
\ref{tab:source} lists the galaxy name, 1950 coordinates, heliocentric
velocity ($cz$), B-band luminosity, luminosity distance, and 
the E+A environment. 
An E+A is identified as lying in a
cluster if it is a member of a cluster in the LCRS group catalog
\citep[e.g.,][]{tuc98} or if there is a known cluster \citep[from the 
NASA/IPAC Extragalactic Database (NED);][]{hel91} at its position and
redshift.  Because the galaxies are at a relatively high redshift ($z
\sim 0.1$) for an {\rm\sc Hi} detection experiment we opted for a
conservative instrumental set up, using the 3 km (C) array as a
compromise between optimizing surface brightness sensitivity and
resolution (15 arcsec). An extended north arm was used in the first
and third session to obtain a rounder beam for the sources at low
elevation. We chose to probe a large velocity range (1400 km
s$^{-1}$), but with a rather poor velocity resolution of typically 50
km s$^{-1}$; this was achieved by using a 6.25 MHz total bandwidth
with 31 spectral channels and online Hanning smoothing.  The
observational parameters are summarized in Table \ref{tab:inst}.

\placetable{tab:source}

\placetable{tab:inst}

We calibrated the data using standard AIPS procedures.  For each field
we made a continuum image covering the full primary beam centered on
the E+A galaxy. We subtracted the continuum from the spectral line
data by making a linear fit to the visibilities over the line-free
channels, which were not affected by the edge effects of the band.
For a first pass we used channels 5 to 27 to make the fit; if {\rm\sc
Hi} was detected, a new fit was made excluding the channels with line
emission.  The search for {\rm\sc Hi} was done in the UV data and
in the image plane.  In the UV plane, we first vector-averaged over
all baselines and over time, which is equivalent to making a
natural-weighted synthesized beam at the field center. We examined the
resulting spectrum for {\rm\sc Hi} line emission. By
varying the weighting of data, we also obtain different spatial resolutions
and, for example, can directly search for extended emission at the
field center. In the image plane, we examined cubes by eye before and
after continuum subtraction. We also smoothed the cubes spatially and
in velocity to search for emission that was either spatially extended
or covered many channels.  All the cubes were made by using natural
weighting to optimize sensitivity.  The rms noise in the full
resolution cubes is typically 0.4 mJy beam$^{-1}$, which is equivalent
to a few $\times$ 10$^{19}$ cm$^{-2}$ in column density sensitivity.
 
\section{RESULTS}

Of the five galaxies, only one was detected in emission --- EA 1.  No
significant {\rm\sc Hi} emission was detected from any of the other
galaxies, nor was radio continuum detected from any of the five
galaxies. In the following sections, we present the {\rm\sc Hi} data
for EA 1 along with the detection limits for the other systems without
{\rm\sc Hi} detections.  The general results are summarized in Table
\ref{tab:results}.

\placetable{tab:results}

\subsection{EA1}

\begin{figure}
\centerline{\psfig{figure=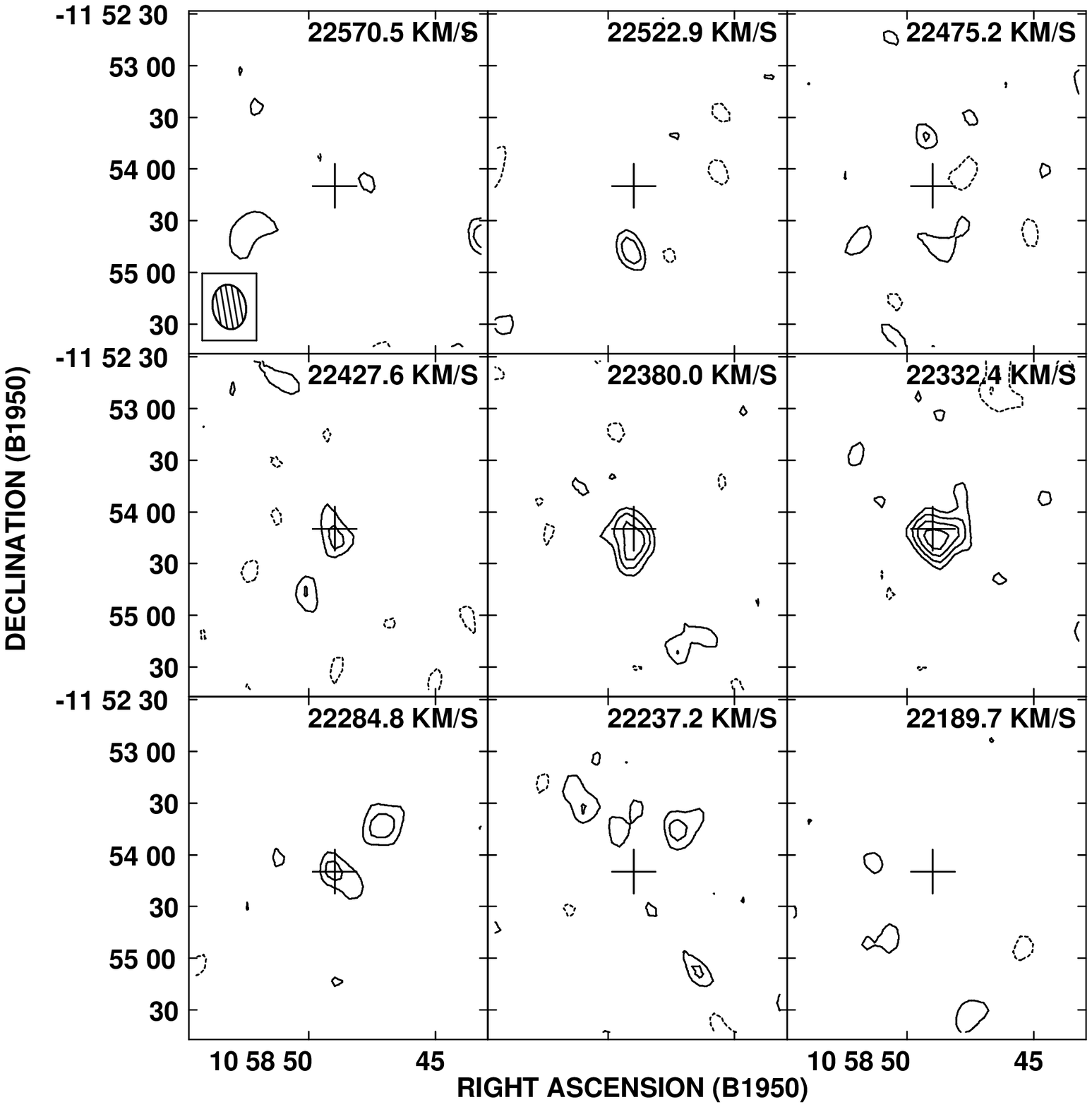,angle=0,height=15cm}}
\caption{\label{fig:kntr}
Channel maps in the region of EA 1. The position of the optical center
is marked by the cross. Contour levels are ($-$4.9, 4.9, 7.3, 9.8,
12.2) $\times$ $10^{19}$ cm$^{-2}$, where $10^{19}$ cm$^{-2}$ = 0.08
mJy per beam. The lowest contour level corresponds to twice the noise
level, while negative contours are denoted by dashed lines. The
synthesized beam is presented by the shaded ellipse on the lower left
of the first panel.}
\end{figure}

\begin{figure}
\centerline{\psfig{figure=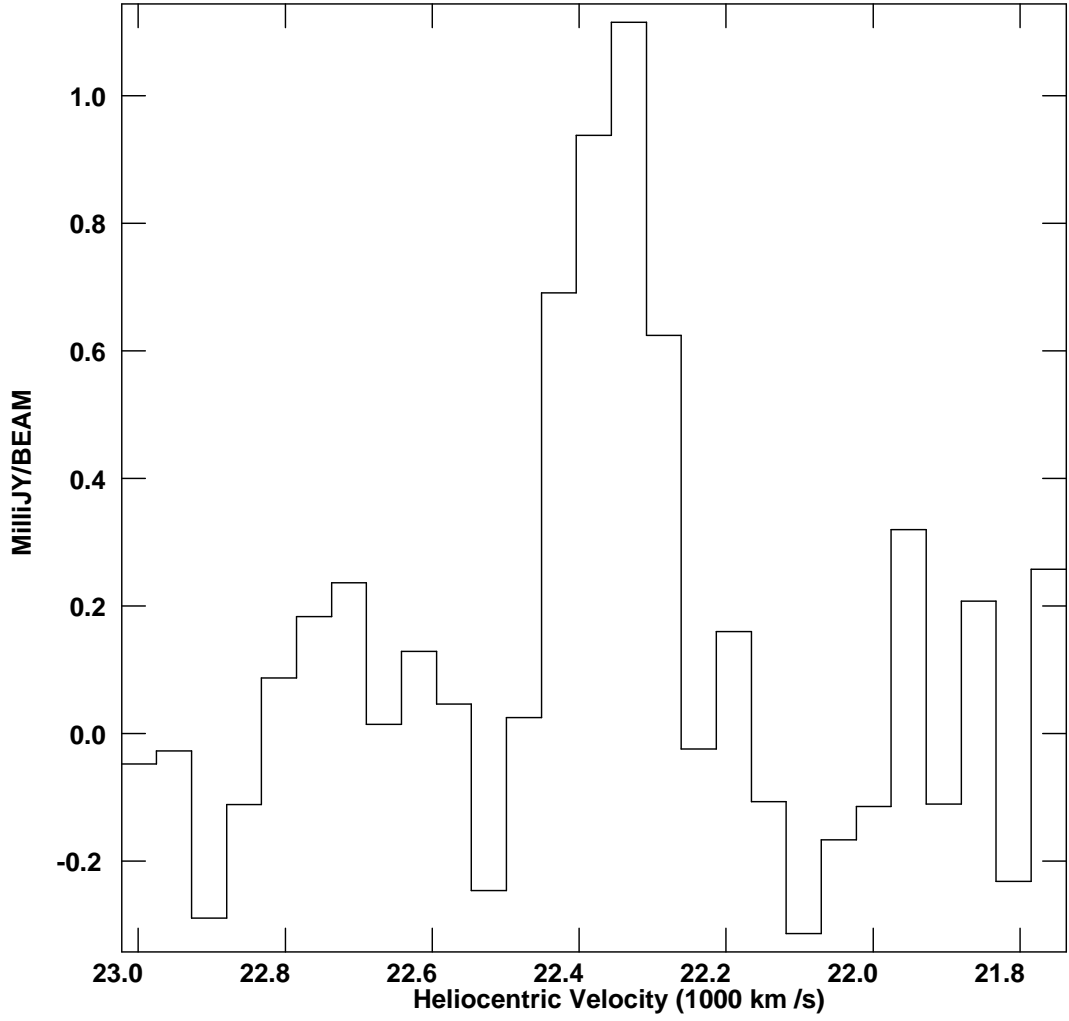,angle=0,height=15cm}}
\caption{\label{fig:ispec}
The {\rm\sc Hi} profile at the center of the galaxy taken from the
image cube.}
\end{figure}

We detect a total {\rm\sc Hi} flux from EA 1 of 0.30 $\pm$ 0.02 Jy km
s$^{-1}$.  We calculate that the mean column density is 1.3 $\times$ $10^{20}$ cm$^{-2}$
and that the total {\rm\sc Hi} mass is 3.5 $\pm$ 0.2 $\times$ $10^{9}$ $h^{-2}$ 
M$_{\odot}$.
In the channel maps, shown in Fig. \ref{fig:kntr}, we see
emission associated with the galaxy over approximately 150 km
s$^{-1}$, from 22284 to 22428 km s$^{-1}$. There is a north-south
velocity gradient in the system with the north side approaching. Near
the central velocity the emission broadens, consistent with the
emission from a moderately inclined, asymmetric disk.  Considering the
very disturbed optical morphology (discussed below) it is also
possible that the {\rm\sc Hi} arises in two kinematically distinct
components that overlap in the central regions.  A profile of the
central {\rm\sc Hi} emission, shown in Fig. \ref{fig:ispec}, also
shows a slight asymmetry.  At the extreme velocities {\rm\sc Hi}
appears northwest and southeast of the galaxy at projected distances
of about 50 $h^{-1}$ kpc. Although this emission is weak, it almost
certainly is real --- it joins smoothly into the main body of emission
in velocity structure and, at very low levels, in spatial distribution
as well. Morphologically and kinematically, the extended emission
appears very similar to the gas-rich tidal tails observed in nearby
mergers \citep[e.g.,][]{hib96}.

\begin{figure}
\centerline{\psfig{figure=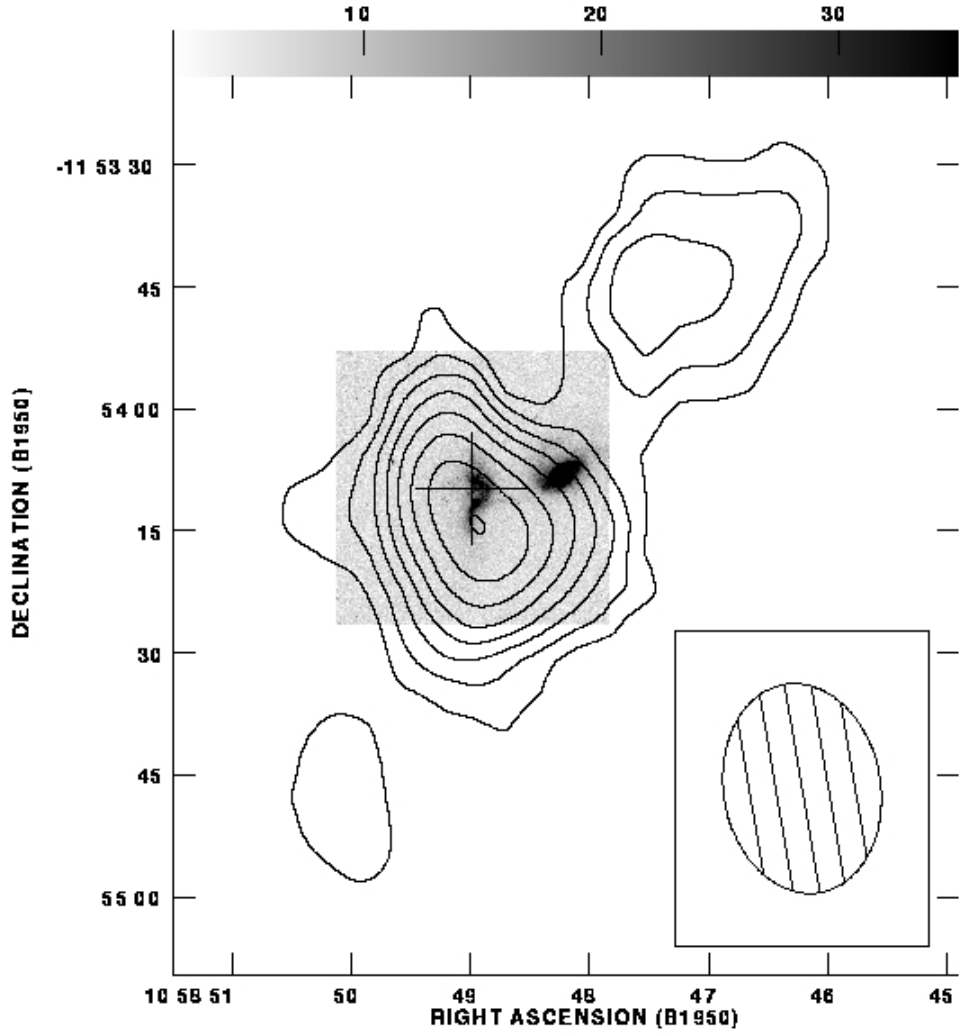,angle=0,height=15cm}}
\caption{\label{fig:hst}
Total {\rm\sc Hi} contour map overlaid on the HST WFPC2 image of EA
1. Contour levels correspond to column densities of $4.9 \times
10^{19}$, $9.8 \times 10^{19}$, $1.5 \times 10^{20}$, $2.0 \times
10^{20}$, $2.4 \times 10^{20}$, $2.9 \times 10^{20}$, $3.4 \times
10^{20}$, and $3.9 \times 10^{20}$ cm$^{-2}$. The cross marks the
position of the optical center of the galaxy. The synthesized beam is
shown on the lower right.}
\end{figure}

\begin{figure}
\centerline{\psfig{figure=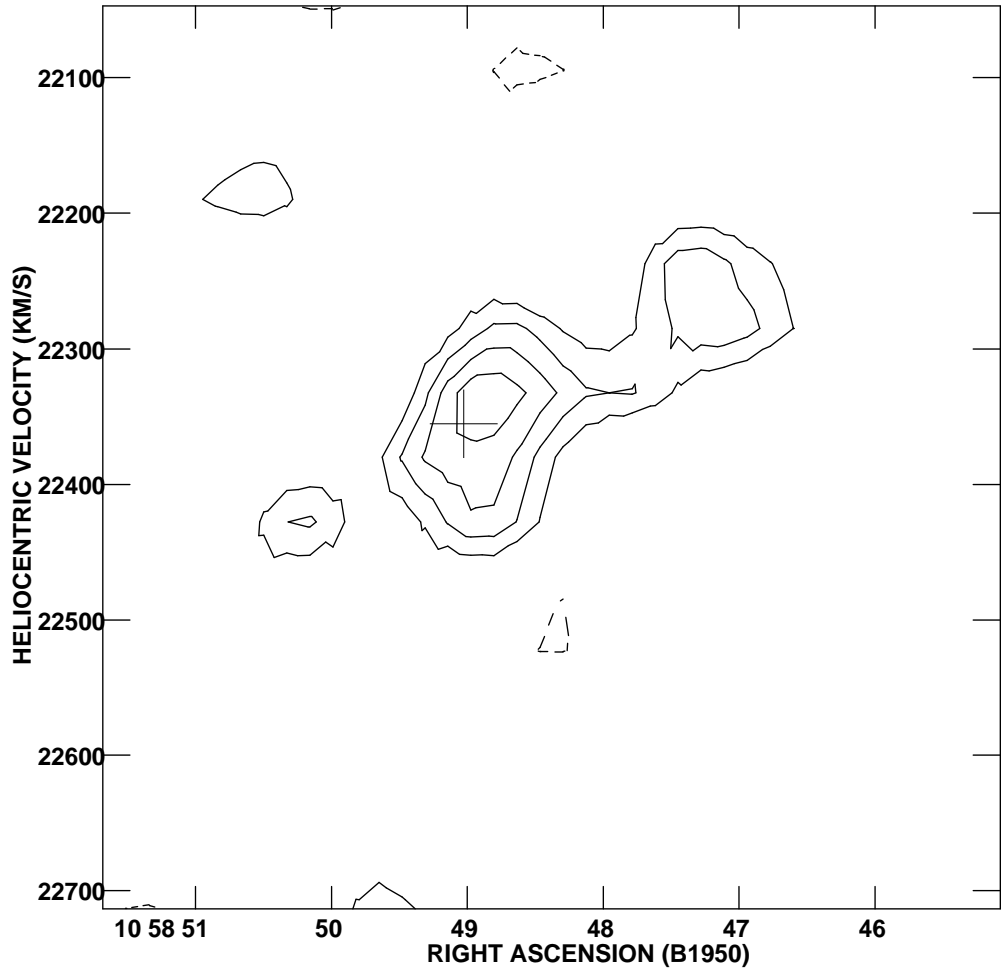,angle=0,height=15cm}}
\caption{\label{fig:pv}
Position-velocity plot of EA 1 along the kinematic major axis. Contour
levels correspond to column densities of $4.9 \times 10^{19}$, $9.8
\times 10^{19}$, $1.5 \times 10^{20}$, and $2.0 \times 10^{20}$
cm$^{-2}$. The cross marks the optical center of EA 1 on this plane.}
\end{figure}

\begin{figure}
\centerline{\psfig{figure=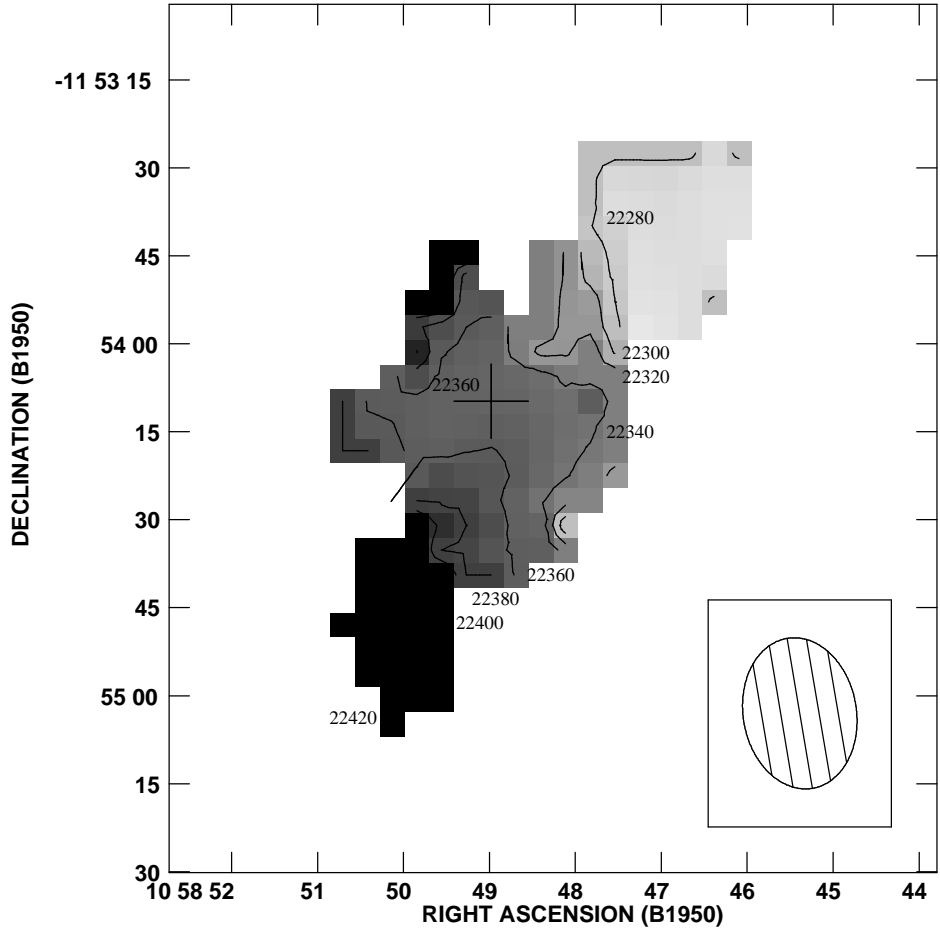,angle=0,height=15cm}}
\caption{\label{fig:vel}
{\rm\sc Hi} intensity-weighted velocity field of EA 1. Numbers denote
{\rm\sc Hi} velocities in km s$^{-1}$. The cross marks the optical
center of the galaxy. }
\end{figure}

In Fig. \ref{fig:hst}, we show the {\rm\sc Hi} emission integrated
over all velocities overlaid on an HST WFPC2 image
(Zabludoff et al, in prep.).  The overlay
further supports the idea that the extended {\rm\sc Hi} is tidal
debris, and the HST image shows that EA 1 looks
very disturbed, much like an ongoing merger.  The HST image also shows
another somewhat disturbed galaxy at a small projected distance;
unfortunately, we do not have a redshift for this galaxy. The only
hint that it may be a true companion is that the {\rm\sc Hi} is
extended in the direction toward the neighboring galaxy. In
Fig. \ref{fig:pv} we show a position-velocity profile along the
kinematical major axis of the {\rm\sc Hi}, approximately the line
connecting the tips of the tidal tails. The asymmetry in the central
{\rm\sc Hi} distribution toward higher velocities is obvious. The
extended {\rm\sc Hi} connects in velocity smoothly with the central
gas, as would be expected from gas that has been tidally
removed. Finally in Fig. \ref{fig:vel} we show the intensity weighted
velocity field of the {\rm\sc Hi} associated with EA 1.

EA1 had already been noted as having tidal features by Z96; however,
it is the combination of the HST image and the VLA {\rm\sc Hi} results
that show how truly strange this system is. The HST image suggests
this is a very young, perhaps ongoing, merger. The two {\rm\sc Hi}
tails suggests that two gas-rich disk galaxies are involved, yet the
current star formation rate as derived from the [O {\rm\sc ii}] is
remarkably low.  From the HST image it is likely that the system is
dusty, hence we may underestimate the amount of [O {\rm\sc ii}]
emission due to extinction. However, a limit to the radio continuum
emission, discussed in \S4, confirms the very low upper limit on the
current star formation rate. We can use the velocities of the
{\rm\sc Hi} tails to give a rough estimate of the time elapsed since
the interaction began \citep{sch82}. Taking $t = 0.8~r_{p}~v^{-1}
\cot{\beta}$, where $r_{p}$ is the projected radius from the center to
the tip of the tail, $v$ is the relative LOS velocity of that tip, and
${\beta}$ is the angle between the radius vector and the line of
sight. The formula assumes that the tail motions are mostly radial, and the
factor 0.8 corrects for minor decelerations and  the non-central tail
origin.  Taking ${\beta}=45^{\circ}$, $r_p= 52 ~h^{-1}$ kpc, and $v=70$ km
s$^{-1}$, we get $t = 6 \times 10^8 ~h^{-1}$ yrs. While it is unclear if the
{\rm\sc Hi} tails are related to the event that caused the starburst,
the derived time scale is consistent with the notion that an
interaction in the last Gyr triggered the original starburst that gave
rise to the E+A spectrum.

\placefigure{fig:kntr}
\placefigure{fig:ispec}
\placefigure{fig:hst}
\placefigure{fig:pv}
\placefigure{fig:vel}

\subsection{The {\rm\sc Hi} Detection Limits}

The smallest {\rm\sc Hi} mass that can be detected anywhere in a
datacube is a 5${\sigma}$ signal per beam per channel.  In Table 
\ref{tab:h1} we list the corresponding {\rm\sc Hi} mass limits, 
using M$_{HI} = 2.35 \times 10^{5} \times d_L^2 \times \frac{\int {\it S
dV}}{1+z} ~h^{-2}$ M$_{\odot}$, where ${\it d_{L}}$ is the luminosity
distance in $h^{-1}$ Mpc, 
$\int {\it S dV}$ is the integrated {\rm\sc Hi} emissivity in Jy km
s$^{-1}$, and $z$ is the redshift of the galaxy.  
These values are valid at the field center.  At larger
distances they have to be corrected for the primary beam response 
because, for example, at a distance of 15$^{\prime}$ from the field center,
these limits are a factor two larger. Our limits are for detections
at random positions in the cubes.

A signal at the position and velocity of an E+A galaxy would be
believable at the 3${\sigma}$ level. Because our only detection, EA1, is
the most nearby of the E+A galaxies, we have also calculated whether
{\rm\sc Hi} emission with the spatial and spectral characteristics of
the gas in EA1 would be detectable at the distances of the other E+A
galaxies.  In the last two columns of Table \ref{tab:h1} we list the
rms noise after smoothing to the velocity extent of EA 1 and the
expected signal per beam at the location of the other galaxies. EA 1
would have been detectable at the 3${\sigma}$ level even at the
greater distances of the other E+A galaxies with the possible
exception of EA 11, for which the signal would only be 2.5${\sigma}$.

\placetable{tab:h1}

\subsection{The Continuum Detection Limits}

In Table \ref{tab:conti} we list the rms noise in the continuum images
and the 3${\sigma}$ detections limits at 1.3 GHz at the location of
the E+A galaxies. The power per synthesized beam is $P_{1.3} = 1.2$
$\times$ $10^{17}$ $\times$ ${\it S}$ $\times$ $d_L^2$,
where $P_{1.3}$ is the power of the source at 1.3 GHz in $h^{-2}$ W
Hz$^{-1}$, ${\it S}$ is the flux in mJy, and $d_{L}$ is the luminosity
distance in $h^{-1}$ Mpc. None of the five E+A's are detected in
continuum emission.  

\placetable{tab:conti}

\section{DISCUSSION}

\subsection{{\rm\sc Hi} Properties} 

Of our five spectroscopically selected E+A galaxies, we have detected
{\rm\sc Hi} in one system, with an {\rm\sc Hi} mass of 3.5 $\times$
10$^9$ $h^{-2}$ M$_{\odot}$, while the other four non-detected systems
have {\rm\sc Hi} upper limits of order 10$^9$ $h^{-2}$ M$_{\odot}$.  What does
this small sample tell us about E+A galaxies as a class? The first
important lesson is that not all E+A galaxies are very gas deficient
--- it is possible to stop star formation without exhausting all the
gas. Secondly, even within our rigorously defined set of E+A galaxies
there is a large spread in {\rm\sc Hi} properties. Although our one
detected galaxy, EA 1, is the bluest galaxy in the entire sample of
Z96, it is not unique in any sense.  EA 1's only distinction is its extremely
disturbed optical morphology --- two components are visible --- but two
galaxies with {\rm\sc
Hi} non-detections, EA 2 and EA 3, are just slightly redder and
are also tidally disturbed. Z96 note
that even with their strict selection criteria, the spectra of the 21
E+A galaxies span a range of spectral characteristics that suggest
that these objects are at different evolutionary stages, have
experienced different evolutionary histories, and/or had
morphologically different progenitors. Can we shed more light on the
evolutionary status of these objects from their {\rm\sc Hi}
properties?
 
The Z96 sample of E+A galaxies is far more extreme than any other
E+A sample in terms of its spectroscopic properties, simply because the 
large number of galaxies ($\sim$11000) in the LCRS allows us to find 
extremely rare objects. Thus
we can only compare the {\rm\sc Hi} properties
of our sample with the little {\rm\sc Hi} data that exist for
previously known, but less well-defined, E+A galaxies in the local
universe, such as the merger remnants of \citet{liu95a,liu95b}
and the cluster galaxies of \citet{cal97}.
The qualitative description of the E+A spectra, in
general, is the strong Balmer absorption lines and the very weak
emission from [O {\rm\sc ii}] and [O {\rm\sc iii}] forbidden lines.
Merger remnants have strong Balmer absorption
lines, but also significant emission from [O {\rm\sc ii}] lines
\citep{liu95a,liu95b,sch96,ken92,car88,bic87}. Several of these galaxies have been
detected in {\rm\sc Hi} emission. For instance, two merger remnants,
NGC 7252, which has clear tidal tails with a total {\rm\sc Hi} mass of
2 $\times$ $10^{9}$ $h^{-2}$ M$_{\odot}$ \citep{hib94}, and
NGC 520, with a total {\rm\sc Hi} mass of 3.9 $\times$ $10^{9}$ $h^{-2}$ M$_{\odot}$,
exhibit strong Balmer absorption lines but have 6 \AA\
[O{\rm\sc ii}] emission line equivalent widths \citep{liu95a}. 
These galaxies have been called E+A galaxies in the
literature, but clearly their spectroscopic properties are different
from those of our E+A sample: contrary to our E+As these galaxies have low
level ongoing star formation.    
Moreover, optically EA1 looks like a much younger 
object than, for example, the merger remnant NGC 7252, and more
similar to the Antennae or NGC 6240, both of which are vigorously
forming stars. This begs the question, why is EA 1's current star
formation rate so low?  Could it be that the gas in EA 1, although
significant in {\rm\sc Hi} mass, is too rarefied to form stars?
Unfortunately, the relatively low spatial resolution of our {\rm\sc
Hi} data makes it impossible to resolve the inner {\rm\sc Hi} mass
distribution in EA 1 and address these questions.  Furthermore, the
upper limits to the {\rm\sc Hi} mass in EA 2 and EA 3 are not low
enough to coin these systems as {\rm\sc Hi} deficient for merging or
interacting systems. That EA 1 is bluer and apparently more gas-rich than EA 2
and EA 3 could point to a difference in gas properties of the
progenitors of these systems or  a different evolutionary history.
On the other hand, EA 1 could represent an early stage of the same evolutionary sequence,
before the gas disperses to the point where it is undetectable in EA 2 and EA 3.

We have measured only upper {\rm\sc Hi} mass limits for the two E+A galaxies,
EA 4 and EA 11, in a cluster environment.
A recent {\rm\sc Hi} study of 11 of the 15 E+A galaxies identified in the 
Coma cluster by \citet{cal93} and \citet{cal97} placed upper limits to the {\rm\sc
Hi} mass of these galaxies of a few $\times$ 10$^7$ $h^{-2}$
M$_{\odot}$ \citep{bra00}.  Because our {\rm\sc Hi} mass  limits are
larger and because the criteria for E+A galaxies
differ between the samples, a direct comparison is not possible.
Observations of larger, well defined, samples are necessary to determine
whether cluster E+A galaxies have less {\rm\sc Hi} than their field counterparts.

The {\rm\sc Hi} upper limits also do not constrain the galaxy type. There
is a large spread in {\rm\sc Hi} content for galaxies of all
morphological types \citep{rob94}, and essentially no type
can be ruled out based on our upper limits.   
Although the {\rm\sc Hi} content of individual galaxies varies
enormously for a given Hubble type, the ratio of {\rm\sc Hi} mass and
blue luminosity is more tightly constrained, log$(M_{HI}/L_{B})$
varies smoothly from $-1$ to just under $0$ from early (SOa) to late
(Sm, Im) types \citep{rob94}.  In Table 4 we list this
ratio for the five galaxies.
The blue luminosities are calculated
from HST PC2 images in the F439W filter (Zabludoff et al., in prep.).
For each galaxy,
we determine the total magnitude within a 100 pixel radius, 
correcting for Galactic extinction and applying a $k$-correction
based on the spectral energy distribution of an A star.  The F439W band
is approximately the Johnson B-band for an A stellar spectrum.
EA 1 stands out with a rather high log$(M_{HI}/L_{B})$ value of $-0.15$;
interestingly, this value is similar to that for tidal tails in gas
rich mergers \citep{hib94}. The upper limits for the other E+A galaxies
fall within the range of values for disk galaxies. 

We also search a volume of $40\arcmin
\times 40\arcmin \times 1400$ km s$^{-1}$ around each of the galaxies.
Taking into account the
primary beam response, the total volume examined down to $5 \times
10^9 ~h^{-2}$ M$_{\odot}$ is $340 ~h^{-3}$ Mpc$^3$.  Using the {\rm\sc
Hi} mass function derived by \citet{zwa97} we expect to detect three
galaxies with $\sim 5 \times 10^9$ $h^{-2}$ M$_{\odot}$ of {\rm\sc Hi} in such
a volume. We only detect one system. This result is somewhat lower than
the expected three, but the difference is perhaps not statistically significant.
However, we note that our fields are not randomly chosen, but are
pointed at known galaxies, and galaxies tend to cluster. Early type
galaxies detected in {\rm\sc Hi} are preferentially found in {\rm\sc
Hi} rich groups \citep{van97}, many late type
(gas-rich) galaxies live in groups, and even  cluster gas-rich
galaxies tend to be subclustered \citep[e.g.,][]{van96}.  The
detection of only one large {\rm\sc Hi} mass in the entire volume
suggests that the environment of the E+A galaxies is mildly {\rm\sc
Hi} deficient.
				    
\subsection{The inferred star formation rate}

The claim that the E+A galaxies have no ongoing star formation is
based on the lack of significant optical emission lines. Our
detection of {\rm\sc Hi} in EA 1 and our upper limits on {\rm\sc Hi} and radio continuum
emission provide an independent way to constrain the on-going star
formation rate.  In this section we compare the rates estimated using 
the measured {\rm\sc Hi} column density, radio continuum emission, and 
[O {\rm\sc ii}] equivalent widths of the E+A galaxies.
The results are listed in Table \ref{tab:sfr}.

\subsubsection{{\rm\sc Hi} Limits}

Although stars form out of molecular gas, {\rm\sc Hi} appears to trace
the star formation rate (SFR) remarkably well \citep{ken89}. Is the detection of a few
$\times$ 10$^9$ $h^{-2}$ M$_{\odot}$ of {\rm\sc Hi} in EA 1 in contradiction
with the claim that no on-going star formation occurs in E+A galaxies?
We can make a crude estimate of the SFR in EA 1 using the Schmidt law,
which relates the gas surface density to the star formation rate by
a simple power law. According to \citet{ken98},

$\Sigma_{SFR}$ = (2.5 $\pm$ 0.7) $\times$ $10^{-4}$ $\Sigma_{gas}^{1.4 \pm 0.15}$  M$_{\odot}$ yr$^{-1}$ kpc$^{-2}$,

\noindent where $\Sigma_{gas}$ is the total (atomic + molecular) gas
surface density averaged over the disks in M$_{\odot}$ pc$^{-2}$.  The
mean {\rm\sc Hi} surface density in the core of EA 1 is 2.2 $\times$
$10^{20}$ cm$^{-2}$, or $\sim$ 1.8 M$_{\odot}$ pc$^{-2}$. If we simply
adopt the Schmidt law to estimate the SFR, taking the mean {\rm\sc Hi}
surface density of the central region of EA 1 as the total gas surface
density and estimating the size of this region to be $\sim$ 940
$h^{-2}$ kpc$^{2}$, we get a SFR of $\sim$ 0.5 $h^{-2}$
M$_{\odot}$ yr$^{-1}$. The upper limits on the {\rm\sc Hi} mass in the
other E+A galaxies provide upper limits on their SFR of the same order
of magnitude (Table \ref{tab:sfr}).

\citet{ski86} points out, however, that there is a surface density
threshold below which no star formation occurs. He finds that the observed local
{\rm\sc Hi} column density threshold for star formation is 1 $\times$
$10^{21}$ cm$^{-2}$, at a resolution of 500 pc.  Although our
resolution is of order 10 kpc, it is still interesting to note that
the {\rm\sc Hi} column density of EA 1 and the column density upper
limits for the other E+A galaxies are well below this threshold.  If
the gas is distributed smoothly over the beam area, then we might
expect no on-going star formation in any of the E+A galaxies even though
ample {\rm\sc Hi} exists.

\subsubsection{Radio Continuum Limits}

Radio continuum emission associated with star-forming
galaxies is thought to be due to synchrotron radiation produced by
relativistic electrons in supernovae remnants. Therefore, the
radio continuum emission provides a probe of massive (M $>$ 8
M$_{\odot}$) star formation in galaxies, independent of dust
obscuration.  \citet{con92} finds a relationship between 1.4 GHz
luminosity and massive SFR of

SFR$_{1.4}$(M $>$ 5 M$_{\odot}$) ~ = ~~$\frac{L_{1.4}}{4.0 \times 10^{21} W
Hz^{-1}}$~ M$_{\odot}$ $yr^{-1}$.

\noindent assuming a Miller-Scalo IMF and using the
Galactic radio luminosity and supernova rate.
We list upper limits to the star formation rate using this formula
in Table 6. 

\citet{sma99} have cast some doubt on the post-starburst
interpretation of the E+A phenomenon after detecting a significant
fraction of cluster E+A galaxies at
z=0.41 in radio continuum.  Radio continuum emission can either be
powered by active galactic nuclei (AGN) or by massive star formation.
Observationally the two can easily be
distinguished by morphology (jet/core morphology vs. extended disk emission).
The radio
power resulting from the two mechanisms is also quite different with a
small range of overlap.  The radio continuum luminosity from AGN is
typically L $\sim$ $10^{23-26}$ $h^{-2}$ W Hz$^{-1}$ at 1.4 GHz, while
that due to star formation is L $\sim$ $10^{21-23}$ $h^{-2}$ W
Hz$^{-1}$ \citep{con89}.  The (unresolved) detections by \citet{sma99}
in CL 0939+4713 at z$=0.41$ have radio luminosities of L $\sim$
$10^{22-23}$ $h^{-2}$ W Hz$^{-1}$ and, based on their radio
luminosity alone, could either be dust-enshrouded starbursts or AGN.
However, these authors suggest that the fact that 4 out of 5 of those 
systems are disk galaxies
favors the starburst interpretation.  Our upper limits of a
few $\times$ $10^{21}$ $h^{-2}$ W Hz$^{-1}$ make it unlikely that our
sample of E+A galaxies host radio loud AGN. Although we cannot rule
out ongoing star formation at a low level (Table \ref{tab:sfr}), our upper limits
are an order of magnitude smaller than the detections by
\citet{sma99}. We do not find evidence for hidden starbursts in the
most extreme local E+A's.
 
\subsubsection{$[${\rm\sc O ii}$]$ Limits}

The [O {\rm\sc ii}] emission line strength is another indicator of on-going
star formation. The derived star formation rates from [O {\rm\sc ii}]
are less accurate than those derived from H$\alpha$, but are
comparable in accuracy to rates derived from the optical continuum
flux and colors \citep{ken92b}. \citet{bar97} find
that if the number of ionizing stars of an {\rm\sc H ii} region is
small ($\sim$ 50), the [O {\rm\sc ii}] equivalent width itself is a good SFR
indicator; however, the method fails if the number of ionizing stars
is much larger than 50 \citep{sta96}. Because of their low current
SFR's, the E+A's are probably in a low ionization regime and the [O
{\rm\sc ii}] equivalent width should provide a rough estimate of the SFR in the E+A galaxies.

The relation between the [O {\rm\sc ii}] luminosity and 
equivalent width is given by \citep{ken92b}

L$_{[{\rm O~II}]} \sim 1.4 \times 10^{29} \times L_{B, \odot} \times
{\rm EW[O ~II]}$ erg s$^{-1}$,

\noindent where $L_{B, \odot}$ is the B band luminosity in
$L_{\odot}$, and EW stands for equivalent width.  The $L_{B, \odot}$
values were derived from our HST data and are listed in Table
\ref{tab:source}. Using the relationship between SFR and 
L$_{[{\rm O ~ II}]}$ \citep{bar97}:

SFR(M$_{\odot}$ yr$^{-1}$) $\sim$ 6.3 $\times$ $10^{-41}$ L$_{[{\rm O~ II}]}$

\noindent we calculate the star formation rates (Table \ref{tab:sfr}) for
all E+A galaxies. A 2$\sigma$ upper limit was used for the line
intensity in the E+A galaxies that were not detected in [O {\rm\sc
ii}].

The different methods used to derive limits on the current SFR in E+A
galaxies each have their own limitations. The estimates based on the
[O {\rm\sc ii}] equivalent width could be affected by
extinction. Although there is a tight relation between radio continuum
and other indicators of SFR (specifically CO emission and FIR flux),
the reason for this correlation is not well understood. The SFR associated with
{\rm\sc Hi} must also depend on the volume density of the gas, which
is not known. Nonetheless, it is reassuring that the low star
formation rates indicated by the [O {\rm\sc ii}] equivalent widths are
not that much lower than the other values. These E+A galaxies are
forming stars at a very modest rate, if at all.

\placetable{tab:sfr}

\section{CONCLUSIONS}

This first search for gas in a rigorously defined set of the most 
extreme local E+A galaxies finds $3.5 \times 10^9 ~h^{-2} ~$M$_{\odot}$ of
{\rm\sc Hi} in one galaxy and obtains upper limits for four
more. Thus not all E+A galaxies are gas deficient; 
E+A formation does not always exhaust the entire gas reservoir.

In the one system detected in {\rm\sc Hi}, the morphology of the gas
supports the hypothesis that the original starburst was induced via a
galaxy interaction. The time elapsed since the interaction started, derived from the {\rm\sc Hi}
kinematics, is roughly consistent with
the time since the starburst must have stopped. However, the
fact that the system contains a substantial amount of {\rm\sc Hi}, and
that the merger appears not yet complete, raises questions of why the
starburst has ceased.

None of the E+A galaxies in our sample were detected in radio
continuum.  The corresponding upper limits on the star formation
rate rule out that these galaxies are highly
obscured, extremely vigorous star forming systems (i.e. $>$ 10
M$_{\odot}$ yr$^{-1}$). Using spectral synthesis models \citep[and
later models]{bru93}, 
one can show that the starbursts that produce 
E+A galaxies formed $\gtrsim 10\%$ of their stellar
mass during a burst of duration $\lesssim 10^9$ yrs; therefore, 
the corresponding SFR during the burst must be $\gtrsim$ 10 M$_{\odot}$
yr$^{-1}$. From a comparison of this SFR and our observational limits, 
we conclude that the SFR in these systems must have
dropped by at least an order of magnitude within the past 1 Gyr.

Although we can exclude large SFR's, none of the available star formation indicators exclude 
star formation rates
at a level comparable to or below that of a weak spiral galaxy.
The [{\rm\sc O ii}] emission or lack thereof implies even lower 
star formation rates than inferred from the {\rm\sc Hi} or radio continuum
luminosities, but the [{\rm\sc O ii}] estimate could be affected by dust.

\acknowledgements
We thank Tod Lauer for providing the B-band luminosities of the 
sample galaxies.
This research was supported by an NSF grant to Columbia University.
AIZ acknowledges support from NASA grant HF-01087.01-96A and HST grant
GO-06835.01-95A. DZ acknowledges support from the David and Lucile
Packard Foundation the Alfred P. Sloan Foundation, and a NASA LTSA
grant (NAG 5-3501).  JCM acknowledges support from an NSF CAREER award
(NSF AST-9876143).  The National Radio Astronomy Observatory is
operated by Associated Universities, Inc., under cooperative agreement
with the National Science Foundation.

\clearpage

\clearpage

\begin{deluxetable}{llllccc}
\tablecaption{Source Parameters 
\label{tab:source}}
\tablewidth{0pt}
\tablehead{
\colhead{Galaxy} & \multicolumn{2}{c}{Coordinate(B1950)} &
\multicolumn{1}{c}{\it cz} & \colhead{ L$_{B}$} & \colhead{ $d_{L}$} &
\colhead{Cluster$?$} \\ 
\colhead{} &  \multicolumn{1}{c}{$\alpha$}  & \multicolumn{1}{c}{$\delta$} & 
\colhead{(km $s^{-1}$ )} & \colhead{(10$^{9}$ L$_{\odot}$)}  & 
\colhead{($h^{-1}$ Mpc)} & \colhead{}}
\startdata
EA 1  &  $10^{h} 58^{m} 48^{s}.98$  &  $-11^{\circ} 54^{\prime} 9\farcs8$
&  $22380 \pm 56$  & 4.9 & 228 & N \\
EA 2  &  $2^{h} 15^{m} 43^{s}.24$   &  $-44^{\circ} 46^{\prime} 36\farcs7$
&  $29600 \pm 42$  & 9.4 & 303 & N \\
EA 3  &  $12^{h} 6^{m} 31^{s}.34$   &  $-12^{\circ} 5^{\prime} 55\farcs4$
&  $24310 \pm 39$  & 16.7 & 248 & N  \\
EA 4  &  $3^{h} 58^{m} 23^{s}.42$   &  $-44^{\circ}  43^{\prime} 40\farcs3$
&  $30350 \pm 30$  & 15.9 & 311 & Y \\
EA 11 &  $1^{h} 12^{m} 34^{s}.57$   &  $-41^{\circ} 38^{\prime} 21\farcs8$
&  $36480 \pm 44$  & 11.4 & 374 & Y \\
\enddata
\end{deluxetable}

\begin{deluxetable}{lllllcc}
\tablecaption{Instrumental Parameters 
\label{tab:inst}}
\tablewidth{0pt}
\tablehead{
\colhead{Galaxy} & \multicolumn{2}{c}{Observing} & \multicolumn{2}{c}{Field
Center(1950)} & \colhead{Array} & \colhead{Velocity} \\
\colhead{} & \multicolumn{1}{c}{Date} & \multicolumn{1}{c}{Time} & 
\multicolumn{1}{c}{$\alpha$} & \multicolumn{1}{c}{$\delta$} & \colhead{} 
& \colhead{km $s^{-1}$}}
\startdata
EA 1 & Mar 1996 & $\sim$ 11 hr & $10^{h} 58^{m} 48^{s}.98$ &
$-11^{\circ} 54^{\prime} 9\farcs8$ & C & 22380 \\
EA 2 & Jun 1997 & $\sim$ 12 hr & $2^{h} 15^{m} 43^{s}.24$ &
$-44^{\circ} 46^{\prime} 36\farcs7$ & CnB & 29600 \\
EA 3 & Mar 1996 & $\sim$ 2 hr & $12^{h} 6^{m} 31^{s}.34$ &
$-12^{\circ} 5^{\prime} 55\farcs4$ & C & 24310 \\
EA 4 & Jan 1996 & $\sim$ 22 hr & $3^{h} 58^{m} 23^{s}.42$ &
$-44^{\circ} 43^{\prime} 40\farcs3$ & CnB & 30350 \\
& \& Jun 1997 & & & & & \\
EA 11 & Jun 1997 & $\sim$ 12 hr & $1^{h} 12^{m} 34^{s}.57$ &
$-41^{\circ} 38^{\prime} 21\farcs8$ & CnB & 36480 \\
 \enddata
\end{deluxetable}

\begin{deluxetable}{lccrc}
\tablecaption{Observation Results 
\label{tab:results}}
\tablewidth{0pt}
\tablehead{
\colhead{Galaxy} & \multicolumn{3}{c}{Synthesized Beam} & 
\colhead{Total {\rm\sc Hi} Flux} \\
\cline{2-4} \\
\colhead{} & \colhead{FWHM} & \colhead{($h^{-1}$ kpc $\times$ $h^{-1}$ kpc)} & \multicolumn{1}{c}{Position angle} & \colhead{(Jy km $s^{-1}$)}}
\startdata
EA 1 & $26\farcs00$ $\times$ $19\farcs70$ & 24.8 $\times$ 18.8 &
$\phantom{00}9.66^{\circ}$ & 0.30 $\pm$ 0.02 \\
EA 2 & $38\farcs05$ $\times$ $16\farcs00$ & $46.3$ $\times$ $19.5$ & 
$\phantom{00}5.64^{\circ}$ & - \\
EA 3 & $42\farcs75$ $\times$ $17\farcs59$ & $43.9$ $\times$ $18.1$ & 
$-41.70^{\circ}$ & - \\
EA 4 & $40\farcs14$ $\times$ $15\farcs32$ & $50.0$ $\times$ $19.1$ & 
$\phantom{0}14.15^{\circ}$ & - \\ 
EA 11 & $34\farcs20$ $\times$ $16\farcs20$ & $49.2$ $\times$ $23.3$ & 
$-15.02^{\circ}$ & -\\
\enddata
\end{deluxetable}

\begin{table}
\caption{The {\rm\sc Hi} Properties \label{tab:h1}}
\tiny
\medskip
\begin{tabular}{lccccccc}
\tableline \tableline
Galaxy & \multicolumn{2}{c}{r.m.s. noise} & Channel Width
& $M_{HI}$ & $\log \left( \frac{M_{HI}}{L_{B,\odot}} \right)$ & 
Observed 200 km $s^{-1}$ r.m.s. &
Intensity expected \\ 
\cline{2-3} \\
& (mJy Beam$^{-1}$) & ($10^{19}$ cm$^{-2}$) & (km
s$^{-1}$) & ($10^{9}$ $h^{-2}$ $M_{\odot}$) &  & (mJy
Beam$^{-1}$) & (mJy Beam$^{-1}$) \\
\tableline
EA 1  & 0.2 & 2.4 & 47.6 & 3.5 $\pm$ 0.2 & $-0.15 \pm 0.07$ & 0.1 & - \\
EA 2  & 0.3 & 3.7 & 49.8 & $<$ 1.5 & $< -0.8$ & 0.2 & 0.8 \\
EA 3  & 0.6 & 5.7 & 48.2 & $<$ 1.9 & $< -0.9$ & 0.3  & 1.2 \\
EA 4  & 0.2 & 2.5 & 50.0 & $<$ 1.0 & $< -1.2$ & 0.1  & 0.7 \\
EA 11 & 0.3 & 4.4 & 51.9 & $<$ 2.3 & $< -0.7$ & 0.2 & 0.5 \\
\tableline
\end{tabular}
\end{table}

\begin{deluxetable}{lcc}
\tablecaption{The Continuum Upper Limits
\label{tab:conti}}
\tablewidth{0pt}
\tablehead{
\colhead{Galaxy} & \colhead{r.m.s. noise} & \colhead{$P_{1.3}$} \\ 
\colhead{} & \colhead{(mJy Beam$^{-1}$)} & \colhead{($10^{21}$ $h^{-2}$ W $Hz^{-1}$)}} 
\startdata
EA 1 & 0.06 & 1.1 \\
EA 2 & 0.1 & 3.3 \\
EA 3 & 0.1 & 2.9 \\ 
EA 4 & 0.07 & 2.4 \\
EA 11 & 0.1 & 7.1 \\
\enddata
\end{deluxetable}

\begin{deluxetable}{lccc}
\tablecaption{Inferred Upper Limit to the SFR ($h^{-2}$ $M_{\odot}$ $yr^{-1}$)
\label{tab:sfr}}
\tablewidth{0pt}
\tablehead{
\colhead{Galaxy} &
\colhead{{\rm\sc Hi}} &
\colhead{Radio Continuum} &
\colhead{[O {\rm\sc ii}]} }
\startdata
EA 1 & 0.5 & 0.3 & 0.1 \\
EA 2 & 0.3 & 0.8 & 0.1 \\
EA 3 & 0.5 & 0.7 & 0.1 \\ 
EA 4 & 0.2 & 0.6 & 0.2 \\
EA 11 & 0.5 & 1.8 & 0.2 \\
\enddata
\end{deluxetable}

\end{document}